\documentstyle[12pt,epsf]{article}
\def\reference{\parskip 0pt\par\noindent\hangindent 0.5 truecm}

\newcommand{\eqb}{\begin{eqnarray}}
\newcommand{\eqe}{\end{eqnarray}}
\newcommand{\gesim}{\,\raisebox{-0.4ex}{$\stackrel{>}{\scriptstyle\sim}$}\,}
\newcommand{\lesim}{\,\raisebox{-0.4ex}{$\stackrel{<}{\scriptstyle\sim}$}\,}
\newcommand{\PSR}{PSR~B1259$-$63}

\newcommand{\Lstar}{L_*}
\newcommand{\Rstar}{R_*}
\newcommand{\Urad}{U_{\rm rad}}
\newcommand{\Lp}{L_{\rm p}}
\newcommand{\epsout}{\epsilon_{\rm out}}
\newcommand{\gammawind}{\gamma_{\rm w}}
\newcommand{\me}{m_{\rm e}}
\newcommand{\rT}{r_{\rm T}}
\newcommand{\diff}{{\rm d}}

\begin{document}
\title{Shock geometry and inverse Compton emission from the wind of a
binary pulsar}

\author{Lewis Ball \and
 Jennifer Dodd}

\date{To appear in {\bf Publ.\ Astronomical Soc.\ Aust.} 18/1\\
Submitted 19 Jul.\ 2000; Revised 8 Nov.\ 2000; Accepted 13 Nov.\ 2000}
\maketitle

{\center
Research Centre for Theoretical Astrophysics,
University of Sydney, N.S.W. 2006, Australia\\
ball@physics.usyd.edu.au}

\begin{abstract}

\PSR\ is a 47ms radio pulsar with a high spin-down luminosity
which is in a close, highly eccentric
3.5-year orbit about a bright stellar companion.
The binary system may be a detectable
source of hard $\gamma$-rays produced by
inverse Compton scattering of photons from the B2e star SS2883
by electrons and positrons in the pulsar wind.
The star provides an enormous density of optical photons
in the vicinity of the pulsar, particularly at epochs near periastron.
We calculate the emission from the unshocked region of
the pulsar wind,
assuming that it terminates at a shock
where it attains pressure balance with the companion's wind.
The spectra and light curves for the inverse Compton emission
from the shock-terminated wind are compared with those for an
unterminated wind.
If the pulsar's wind is weaker than that from the companion star,
the termination of the wind decreases the inverse Compton flux,
particularly near periastron.
The termination shock geometry has the effect of decreasing the
asymmetry of the $\gamma$-ray light curve around periastron,
which arises because of the asymmetrical variation of the
scattering angle.
\end{abstract}

\bigskip

\section{Introduction}

\PSR\ is one of only three known radio pulsars which have a main
sequence star binary companion.
Such systems provide a unique environment for inverse Compton scattering
because of the presence of an enormous density of low energy (optical)
photons which serve as targets for electrons and positrons
in the pulsar wind.
The modest spin-down luminosity, and relatively high distance to two
of these systems together imply that they are unlikely to be detectable
sources of inverse Compton emission.
In contrast, \PSR\ is a galactic radio pulsar with a high spin-down
luminosity ($\Lp = 8.3\times 10^{28}\;$W) and is relatively nearby
($1.5\,$kpc).
If just 0.1\% of the wind luminosity is
scattered into hard $\gamma$-ray photons the resulting flux
should be detectable using current $\gamma$-ray telescopes.

\PSR\ is in a highly eccentric orbit ($e\sim 0.87$) around
SS2883, a B2e star of radius $\Rstar\sim 6 R_\odot$ and
luminosity $\Lstar\sim 8.8\times 10^3\,L_\odot$
[Johnston et al.\ 1992, 1994, 1996].
At periastron the pulsar is only $23\Rstar\approx 10^{11}\,{\rm m}$
from its companion.
Around periastron there are $3.4\times 10^{11}\,\rm cm^{-3}$ Be-star
photons at the pulsar,
corresponding to an energy density of
$\Urad\sim 6\times 10^{11}\,\rm eV\,cm^{-3}$.
This is some 11 orders of magnitude larger
than the typical background target density available for
inverse Compton scattering by the winds of isolated pulsars.

The Crab is presently the only pulsar for which
observations place any constraints on the physical parameters
of a pulsar wind.
Observations of the \PSR\ system aimed at detecting hard $\gamma$-ray
inverse Compton emission are planned during 2000.
If successful they should provide the first direct
probe of the freely-expanding region of the wind
of any rotation-powered pulsar,
and serve to constrain wind models and parameters.

In the absence of an efficient deceleration process
the relativistic wind of a pulsar will expand freely until it attains
pressure balance with the surrounding medium.
If the wind is still supersonic at the point of pressure balance it
will be bounded by a termination shock.
Kirk, Ball \& Skj\ae raasen [1999]
investigated the $\gamma$-ray emission
from \PSR\ resulting from inverse Compton scattering by
the shocked pulsar wind, downstream of the termination shock,
where the pulsar wind electrons and positrons have been accelerated and
isotropised as the radial flow of the wind is disrupted.
The effects of the cooling of the shocked wind by inverse Compton
scattering were included by Tavani \& Arons [1997] in their model for
the synchrotron emission from the system.
Ball \& Kirk [2000a] considered the effects of
inverse Compton scattering by the unshocked pulsar wind,
upstream of the termination shock.
They calculated the deceleration of the wind that occurs
as a result of `inverse Compton drag' as energy and momentum are
transferred from the wind to the scattered photons.
In particular it was shown that the inverse Compton losses were
unlikely to contain the wind of \PSR\ before it attains the radius
at which pressure balance with the Be-star outflow is likely to occur.
Ball \& Kirk [2000a] then calculated the emission from the
freely-expanding wind subject to the assumption that the wind
was not terminated by pressure balance.

In this paper we recalculate the inverse Compton emission from
the unshocked wind of \linebreak\PSR\ including the effects of
termination of the radially-expanding pulsar wind as a result of
pressure balance with the Be-star wind.
The relative strengths of the pulsar and Be-star winds
are unknown.
If the Be-star wind dominates then the termination shock
will be close to, and wrapped around, the pulsar.
This will have the effect of decreasing the 
emission from the unshocked portion of the wind,
and the effects vary over the binary orbit
of the system.
We hope that inverse Compton $\gamma$-rays will be detected from
\PSR\ in the near future,
providing data that can be compared with such models.

The calculation of the geometry of the termination shock
is outlined in \S 2.
In \S 3 we present $\gamma$-ray spectra and light curves
from the terminated wind,
and compare them with those calculated for the unterminated wind
by Ball \& Kirk [2000a].
Our conclusions are presented in \S 4.

\section{Shock geometry}

The winds of the pulsar and its companion star will generally be
separated by a pair of termination shocks separated by a contact
discontinuity.
The shock structures enclose the star with the weaker wind.
Giuliani [1982] developed a general description of
axisymmetric flows that can be applied to the interaction of two
such winds.
Girard \& Willson [1987; hereafter GW87] followed this description
and derived the equations that describe the shape
and position of the boundary separating two stellar winds
subject to the following assumptions:
\begin{enumerate}
\itemsep=0pt
\parsep=0pt
	\item The shape of the shock is not affected by the
	      orbital motion of the two stars.
	      This implies azimuthal symmetry about the line
	      joining the two stars.
	\item The shock region has negligible thickness.
	\item Both winds are emitted radially with no angular
	      dependence.
	\item The winds are non-relativistic.
\end{enumerate}
The resulting set of three ordinary differential equations (ODEs)
can be integrated directly to find the position of the boundary,
which can be loosely referred to as the `shock',
given the assumption of negligible thickness.
The winds are characterised by their mass loss rate $\dot M$
and (constant) radial velocity $u$ and these enter the equations
that describe the shock geometry only in terms of the ratios
\eqb
m={\dot{M_1}\over \dot{M_2}} \qquad {\rm and} \qquad w={u_1\over u_2}
\label{mw}
\eqe
where the subscripts refer to the different stars.
GW87 argued that the results obtained depended
only on the quantity $\eta = m w$,
although $m$ and $w$ enter the ODEs separately.
In an alternative formulation presented by Huang \& Weigert [1982]
the quantities $m$ and $w$ only enter the ODEs as their product $\eta$.

Assumption (4) listed above is almost certainly not valid for a
pulsar wind, which is likely to be highly relativistic.
Modelling of the nebula produced by the wind of the Crab pulsar
implies that the wind has a bulk Lorentz factor of $\gamma\sim10^6$
at radii well beyond the light cylinder
[Kennel \& Coroniti 1984].
Melatos, Johnston \& Melrose [1995] considered the interaction of
a relativistic wind of \PSR\ comprised primarily of electrons
and positrons,
with a wind from SS2883 dominated by much more massive ions.
They showed that when the ions are dominant in both
number and energy density, as is expected in \PSR/SS2883,
the dependence of the geometry only on the product of $m$ and $w$ is lost.
Nevertheless, for a given value of $\eta$ the opening angles
calculated by Melatos, Johnston \& Melrose [1995] are within a
factor of $\sim 2$ of the results of
GW87 for a wide range of values of $w/m$.

In the following we approximate the shock position in the
\PSR/SS2883 system using the formulation of GW87.
This minimises the number of poorly known wind parameters that
affect the results.
The shock position is represented by its radial distance from the
pulsar $\rT(\alpha)$ where $\alpha$ is the angle between the
lines joining the pulsar to the shock and the pulsar to the Be star.

%
%
\begin{figure}[!hbt]
\epsfxsize=8 cm
\centerline{\epsffile{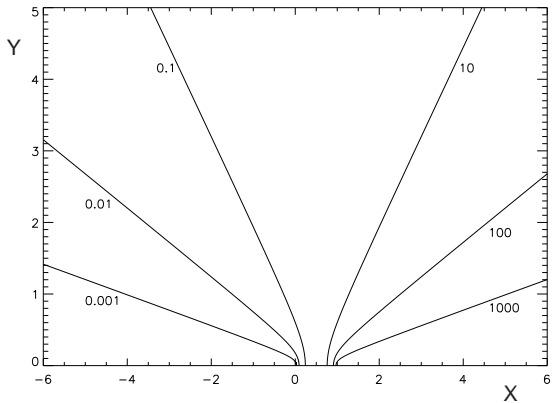}}
\caption{Shock positions calculated from the formulation of GW87.
The pulsar is at (0,0) and the Be star is at (1,0).
The values of $\eta$ are as labelled.
}
\vspace{0.5cm}
\label{shock}
\end{figure}

Figure 1 shows the position of the termination shock (TS)
obtained from numerical solution
of the ODEs of GW87 for six different values of $\eta$.
The quantity $\eta$ is the ratio of the rates at which the stars are
transferring momentum to their winds.
When star 1 is taken to be the pulsar and star 2 is the Be star it
follows that
\begin{equation}
\eta=\frac{L_p/c}{\dot M v}
\label{eta}
\end{equation}
where $\dot M$ and $v$ are the mass loss rate and speed
of the Be-star wind,
and $\Lp$ is the spin-down luminosity of the pulsar.
When $\eta < 1$ the Be-star wind dominates the pulsar wind,
and the TS wraps around the pulsar.
As $\eta$ increases the apex of the TS moves further from the pulsar,
the opening angle increases, and for $\eta > 1$ the pulsar
wind is dominant and the TS wraps around the Be star.

\begin{figure}[!hbt]
\epsfxsize=6 cm
\centerline{\epsffile{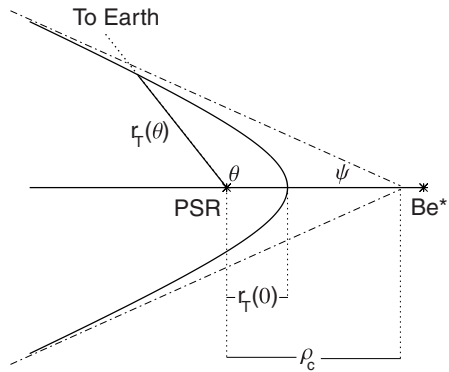}}
\caption{Geometry of the termination shock.
}
\vspace{0.5cm}
\label{geom}
\end{figure}

The standoff distance from the pulsar to the apex of the TS is
\eqb
\rT(0) = \frac{\sqrt{\eta}}{1+\sqrt{\eta}}\; D
\label{standoff}
\eqe
where $D$ is the stellar separation.
At large distances from the stars the momenta of the winds are
almost parallel and the shock tends asymptotically to a cone
characterised by a half-opening angle $\psi$
and distance to the apex $\rho_{\rm c}$.
A useful empirical approximation for $\psi$ as a function of $\eta$
is [Eichler \& Usov 1993]:
\eqb
\psi=2.1\left (1-\frac{\bar{\eta}^{\frac{2}{5}}}{4}\right )
\bar{\eta}^{\frac{1}{3}}
\label{psi}
\eqe
where $\bar{\eta}=\min(\eta,\eta^{-1})$.
Figure \ref{geom} shows the geometry of the termination shock and
illustrates the parameters used to describe it.
We use $\theta$ to denote the value of $\alpha$ corresponding to the
line of sight from Earth to the pulsar;
i.e.\ $\theta$ is the angle between the line of sight
and the line from the pulsar to the Be star.

%

The Be star companion of \PSR\ possesses an excretion disk
which the pulsar passes through near periastron
[Johnston et al.\ 1996, 1999; Ball et al.\ 1999].
This disk will dominate the termination of the pulsar wind
for 50 days or so near periastron,
greatly complicating the physics
of the pulsar wind system at these epochs.
Its effects on unpulsed X-ray emission close to
periastron have been considered by Tavani \& Arons [1997].
However its presence is irrelevant for the majority of the
1237 day binary orbit and it is neglected here.
The results presented below can therefore not be taken to be
accurate at epochs close to periastron.

\section{Inverse Compton scattering}

Ball \& Kirk [2000a; Equation 15] derived an expression for the
radiation transfer that determines the 
intensity of the radiation emitted at the normalised energy
$\epsilon_{\rm out}=E_{\rm out}/\me c^2$.
Furthermore, it was shown that absorption of the scattered
photons due to pair production on the Be-star photons is negligible.
It follows that when the termination of the wind is included,
the scattered intensity can be written as a line of sight integral
of the scattering due to electrons moving radially out from the pulsar,
\eqb
I(\epsilon_{\rm out})=\int_{0}^{\rT(\theta)}
\left({\Lp\over4\pi}\right){\epsout\over \gammawind(0)\me c^3} 
\;
\frac{{\rm d}N_{\gamma}}
{{\rm d}\epsilon_{\rm out}{\rm d}t}\,{\rm d}s \;,
\label{intensity}
\eqe
where the fraction of the wind momentum carried by ions is
assumed to be negligible.
The quantity $\gammawind(0)$ is the initial Lorentz factor of the
pulsar wind and ${\rm d}N_{\gamma}/{\rm d}\epsilon_{\rm out}{\rm d}t$
is the differential rate of emission of inverse Compton scattered photons
by a single electron.
In the calculations of Ball \& Kirk [2000a] the TS was
assumed to be very distant from the pulsar, so that the upper
limit of the integration was effectively infinity.

\begin{figure}[!hbt]
\epsfxsize=6 cm
\centerline{\epsffile{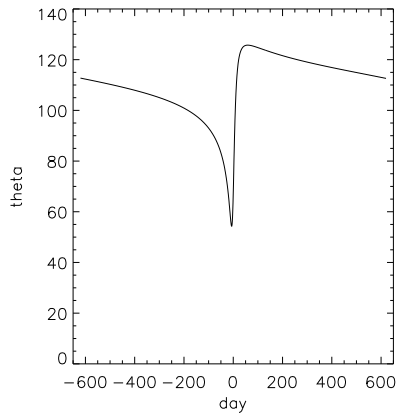}}
\caption{The variation of the angle $\theta$
between the line of sight to \PSR\
and the line joining the pulsar and its Be-star companion
over the binary orbit.
Day 0 corresponds to periastron, which next occurs on 2000 October 17.}
\label{theta}
\end{figure}

The termination of the pulsar wind will necessarily have the largest
effect on its inverse Compton emission for small values of $\theta$,
since regardless of the value of $\eta$, $\rT(\alpha)$ is a
monotonically increasing function.
Furthermore, these effects will be most significant if $\eta$ is small,
since in such cases the pulsar wind terminates close to the pulsar.
The orbit of the \PSR\ system is inclined at $i=35^\circ$ to the
plane of the sky, so an observer samples angles $\theta$
between $90-i=55^\circ$ and $90+i=125^\circ$ over the binary period.
The variation of $\theta$ over the orbit is shown in Figure \ref{theta}.
If $\eta \ll 1$ the pulsar wind will terminate
very close to the pulsar for small $\theta$,
decreasing the observable emission from the unshocked wind
most significantly at binary phases which correspond to values of
$\theta$ that are close to the minimum,
i.e.\ between around day $-200$ and periastron.

%
%
%

\subsection{Spectra}

%
%
\begin{figure}[!hbt]
\epsfxsize=12 cm
\centerline{\epsffile{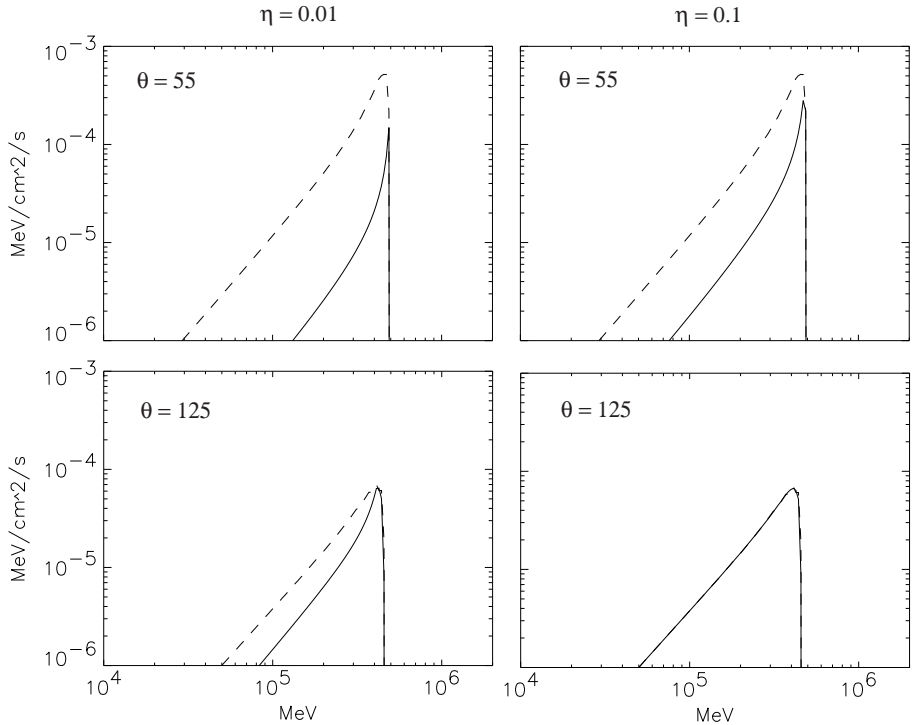}}
\caption{Inverse Compton emission spectra
for the unterminated (dashed lines)
and terminated (solid lines) pulsar wind,
for $\eta=0.01$ (left panels) and $\eta=0.1$ (right panels).
The top row shows the spectra for $\theta=55^{\circ}$,
and the bottom row is for $\theta=125^{\circ}$,
the extremes of this angle that can be sampled from Earth.
All the spectra shown have been calculated for a pulsar wind
with an initial Lorentz factor $\gammawind(0)=10^6$.
}
\label{spectra}
\end{figure}

Figure \ref{spectra} shows the spectra of inverse Compton emission
for two values of $\eta$ for which the Be-star wind dominates the
pulsar wind, which therefore terminates quite close to the pulsar
for small $\theta$.
The quantity plotted is the spectral energy distribution $EF_E$ 
where $F_E$ is the energy flux at Earth of
inverse Compton scattered photons of energy $E$.
The solid lines show the results when the
termination of the pulsar wind is included,
and the dashed lines show the spectra calculated by Ball \& Kirk [2000a]
when the shock termination is very distant from the pulsar.
The spectra are shown assuming that the separation of the pulsar and the
Be star is that which applies at periastron,
and have been calculated for the extreme values of
$\theta$ that can be sampled from Earth.
Both of these extreme values occur very close to periastron,
as can be seen from Figure \ref{theta},
so the assumption of periastron separation is appropriate.
Furthermore, it is assumed that the spectrum of target photons
from the Be star can be approximated as a monochromatic
distribution at a normalised energy
$\epsilon_0 = 2.7k_{\rm B}T_{\rm eff}/(m c^2)\approx 10^{-5}$
where $T_{\rm eff}$ is the star's effective temperature.
Calculations including a
more realistic Be-star spectrum have been presented by
Ball \& Kirk [2000b] and Kirk, Ball \& Skj\ae raasen [2000].
The effects of including a better approximation to the
target photon spectrum are generally small, but may warrant closer
investigation if the system is detected in hard $\gamma$-rays.

All the spectra in Figure \ref{spectra} are sharply peaked
at an energy close to
$5\times 10^5\;\rm MeV$, somewhat lower than the Thomson
limit of $\gammawind(0)^2 \me c^2 \epsilon_0\approx5\times 10^6\;\rm MeV$
because of the importance of Klein--Nishina effects.
In the Klein--Nishina regime, i.e.\ when $\gamma\epsilon_0 \gesim 1$,
a scattering particle loses a significant fraction of its
energy to the scattered photon and the maximum scattered energy
is then $\epsilon_{\rm out}^+\sim \gamma$.
The upper limit of the emission spectrum is unaffected by the
inclusion of the TS,
since the scatterings that produce photons of this energy
involve undecelerated electrons
(with Lorentz factors $\sim \gammawind(0)$)
and occur very close to the pulsar.
As the pulsar wind propagates away from the pulsar the wind
electrons lose energy via the inverse Compton scattering process.
Lower energy scattered photons are generated farther from the
pulsar, and so tend to be suppressed by the termination of the wind.

For small $\theta$ and small $\eta$ the TS is
very close to the pulsar where inverse Compton scattering
is still very efficient because the target photon density is large
and the scatterings are close to head on.
The wind is terminated before it has been
decelerated to Lorentz factors significantly below $\gammawind(0)$,
so the scattering electrons are essentially monochromatic,
as are the target photons in our approximation.
This has the effect of dramatically reducing the emission
at all energies below the upper cutoff,
producing a scattered spectrum that is even more sharply peaked
than that resulting from the unterminated wind.
Such a case is illustrated in the top left panel for $\eta=0.01$
and $\theta=55^\circ$, for which the scattered spectrum is
essentially monochromatic at the upper cutoff.

The TS radius increases rapidly with increasing
$\alpha$ (for a given $\eta$), with the rate of increase approaching
infinity as $\alpha$ approaches $180^\circ-\psi$ where $\psi$ is
the half-opening angle of the cone and $\eta<1$ is assumed.
At energies near the upper cutoff the scattered spectrum from the
terminated wind therefore rapidly approaches that from the
unterminated wind as $\theta$ increases.
The suppression at lower energies,
which are produced by scatterings well away from the pulsar where the
wind has been significantly decelerated,
persists to larger $\theta$.
This effect is dramatically illustrated by a comparison between each
top panel of Figure \ref{spectra} with the corresponding lower panel.
For the case where $\theta=125^\circ$ and $\eta=0.01$
the emission at and just below the cutoff energy is essentially unaffected
by the inclusion of the TS,
because the scatterings that produce photons of these energies occur
closer to the pulsar than does the TS.

For larger values of $\eta$
the TS occurs further from the pulsar.
It follows that for a given $\theta$,
the difference between the inverse Compton emission from the
terminated and the unterminated wind decreases as $\eta$ increases.
For the case shown in the lower right panel of Figure \ref{spectra},
with $\eta=0.1$ and $\theta=125^\circ$, the TS occurs
so far from the pulsar that it has essentially no effect on
the scattered emission in the energy range shown,
though it does reduce the scattered flux at even lower energies.

\subsection{Light curves}

For $\eta>1$ the TS wraps around the Be star
and the line of sight to the pulsar does not
intersect the shock if $\theta > \psi$.
Equation (\ref{psi}) implies that for $\eta \gesim 7$,
$\psi \lesim 55^\circ$ which is the minimum observable value of $\theta$
over the binary orbit of \PSR.
Thus if $\eta \gesim 7$ the line of sight doesn't intersect
the TS at any phase of the binary orbit,
the TS
has no effect on the observable inverse Compton emission from
the freely-expanding wind,
and the results of Ball \& Kirk [2000a] require no modification.

For $\eta<1$ the TS wraps around the pulsar
and is intersected by the line of sight to the pulsar if
$\theta < 180^\circ -\psi$.
It follows from Equation (\ref{psi}) that if $\eta \lesim 0.14$,
$180^\circ - \psi \gesim 125^\circ$
which is the maximum observable value of $\theta$,
and thus the line of sight to the pulsar intersects the
TS at all binary phases.

%
%
\begin{figure}[!hbt]
\epsfxsize=12 cm
\centerline{\epsffile{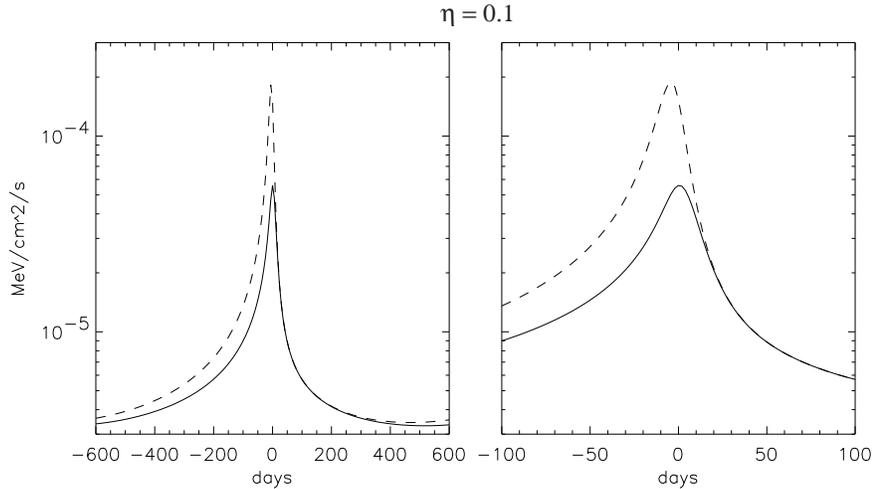}}
\caption{Light curves showing the integrated energy flux at the Earth from
\PSR\ over the whole orbital period (left panel),
and over 200 days centred on periastron (right panel).
The initial Lorentz factor of the pulsar wind is $\gammawind(0)=10^6$.
The dashed curves show the emission from an unterminated wind
as calculated by Ball \& Kirk [2000a].
The solid curves show the emission when the wind is terminated
by a shock whose position is determined by
the parameter $\eta=0.1$.}
\label{light.1}
\end{figure}

Figure \ref{light.1} shows the orbital variation of the integrated
flux density $\int F_E \,\diff E$,
expected from \PSR\ for $\gammawind(0)=10^6$.
The dashed curve shows the emission from the unterminated wind
as calculated by Ball \& Kirk [2000a].
The solid line shows the emission when the wind
is terminated by a shock at the position shown in Figure \ref{shock}
for $\eta=0.1$.
The TS is wrapped around the pulsar and is
intersected by the line of sight to the pulsar at all
binary phases.
However, between days $+50$ and $+400$ the intersection point
is sufficiently distant from the pulsar that it is beyond those
radii where the majority of the inverse Compton scattering occurs.
The termination of the wind therefore has little effect in reducing the
inverse Compton emission in this period,
and the light curve from the terminated wind
is indistinguishable from that of an unconfined wind.

The effect of the termination of the wind is asymmetric because
of the asymmetry in the dependence of the line-of-sight
angle $\theta$ about periastron.
The angle $\theta$ decreases very rapidly towards its minimum
value just prior to periastron,
which implies that the radius at which the line of sight to the pulsar
intersects the TS does the same.
The increase from the minimum value of $\theta$ to the maximum then
occurs very rapidly, starting just a few days before periastron.
The termination of the wind therefore decreases the emission from the
freely expanding portion of the pulsar wind most effectively 
prior to periastron,
and thus has the effect of decreasing the characteristic asymmetry
in the $\gamma$-ray light curve about periastron.
For $\eta=0.1$ Figure \ref{light.1} shows that the effect of the
TS is to reduce the maximum integrated inverse Compton
flux density by a factor of $\sim4$,
to make the time at which the maximum occurs a few days later,
and to reduce the asymmetry about periastron measured by the ratio of
the integrated fluxes at days $\pm50$, from $\sim 3$ to $\sim 1.6$.

%
%
\begin{figure}[!hbt]
\epsfxsize=12 cm
\centerline{\epsffile{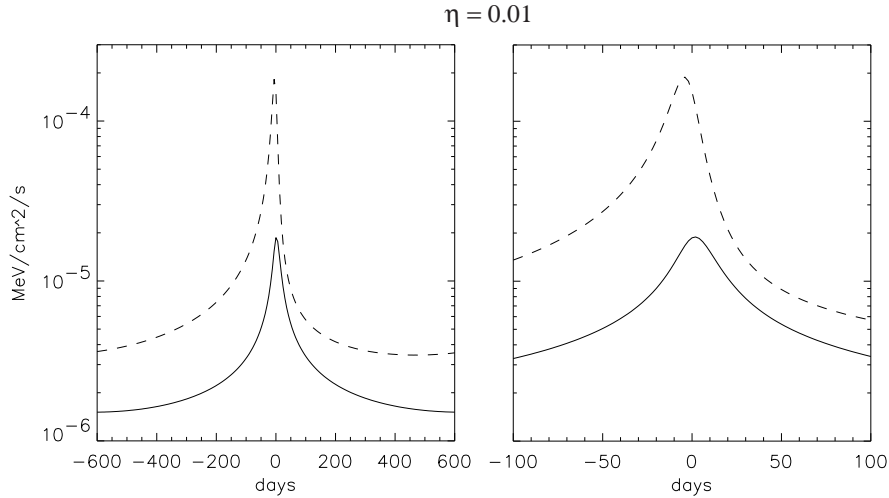}}
\caption{Light curves showing the integrated energy flux at the Earth from
\PSR\ when the wind 
is terminated by a shock whose position is determined by
the parameter $\eta=0.01$ (solid curve).
The initial Lorentz factor of the wind is $\gammawind(0)=10^6$.
The dashed curves show the emission from an unterminated wind
as calculated by Ball \& Kirk [2000a]
(and are identical to those in Figure \ref{light.1}).}
\label{light.01}
\end{figure}

Figure \ref{light.01} shows the light curve for a shock-terminated wind
with $\eta=0.01$.
In this case the shock is so tightly wrapped around the pulsar
that its intersection with the line of sight to the pulsar
always occurs at a radius where inverse Compton scattering is still
effective.
The emission from the terminated wind is thus substantially
less than that from an unconfined wind throughout the binary orbit.
The observable flux in this case is reduced by a factor of between
1.5 and 3 over roughly 80\% of the orbital period.
The maximum integrated hard $\gamma$-ray flux density
is a factor of $\sim 10$ lower than that from an unterminated wind
and occurs $\sim 6$ days later, and the $\pm50$ day asymmetry is actually
reversed to $\sim0.95$.

\section{Discussion and conclusions}

Ball \& Kirk [2000a] argued that inverse Compton emission from the
freely-expanding portion of the wind of \PSR\ should be detectable
at energies somewhere between GeV and TeV,
depending on the wind parameters.
The results presented in Section 3 suggest that even if the wind
of the Be star dominates that of the pulsar, terminating it
in a shock which wraps around the pulsar,
the scattered emission should still be detectable for a wide
range of wind parameters.
However, the TS has the effect of decreasing the peak
integrated $\gamma$-ray flux from the freely-expanding wind,
and of decreasing the asymmetry which
otherwise produces a characteristic light curve which shows
a higher $\gamma$-ray flux before periastron than after.

Even if the Be-star wind dominates to such a degree that
the unshocked pulsar wind is not detectable,
the models of Kirk, Ball \& Skj\ae raasen [1999]
indicate that inverse Compton emission from the shocked
pulsar wind should be above the thresholds of current detectors.

The scattered $\gamma$-ray signal from this system is most likely
to comprise a combination of emission from the shocked and unshocked
regions of the pulsar wind.
The reduced asymmetry and periastron to apastron ratio
of the emission from the terminated wind may make it harder
to deconvolve the two contributions,
but only detailed modelling of real data will provide such answers.

The best opportunity for detecting such emission in the near future
is afforded by the new CANGAROO II imaging Cherenkov telescope located
in Australia [Yoshikoshi et al.\ 1999].
When the present lack of operational $\gamma$-ray telescopes
is relieved by the launch of the GLAST and INTEGRAL observatories,
inverse Compton emission from this pulsar system may well be detectable
at lower energies in the MeV--GeV range.

The next periastron of \PSR\ occurs in October 2000.
CANGAROO II observations of the system will not be possible at times
near periastron because at that time the object will be too
close to the Sun.
In any event, the light curves shown in
Figures \ref{light.1} and \ref{light.01} are not realistic between
about days $-25$ and $+25$ because we have not modelled the interaction
with the Be-star disk which is important at such epochs.
Observations of the \PSR\ system with CANGAROO II are planned for
July and December 2000, when our models of the inverse Compton emission
from the system should be applicable.
If such observations are successful in detecting hard $\gamma$-ray
emission from this unique pulsar system they should provide valuable
insights into the properties of the pulsar wind.
In particular, detection of inverse Compton $\gamma$-ray emission
from the system should help to constrain values of the
bulk Lorentz factor of the pulsar wind and
the fraction of the spin-down luminosity carried by
electrons and positrons compared to ions.

\section*{References}

\reference
Ball, L., \& Kirk, J. G., 2000a,
Astropart.\ Phys., 12, 335

\reference
Ball, L., \& Kirk, J. G., 2000b,
ASP Conf., 202, 531

\reference
Ball, L., Melatos, A. M., Johnston, S., Skj\ae raasen, O., 1999,
ApJ, 541, L39

\reference
Eichler, D., Usov, V., 1993,
ApJ, 402, 271 

\reference
Girard, T., Willson, L. A., 1987,
A\& A, 183, 247 

\reference
Giuliani, J. L., 1982,
ApJ, 256, 634 

\reference
Huang, R. Q., Weigert, A., 1982,
A\&A, 112, 281

\reference
Johnston, S., Manchester, R. N., Lyne, A. G., Bailes, M., Kaspi, V. M.,
Qiao, G., D'Amico, N., 1992,
ApJ, 387, L37

\reference
Johnston, S., Manchester, R. N., Lyne, A. G.,
Nicastro, L., Spyromilio, J., 1994,
MNRAS, 268, 430

\reference
Johnston, S., Manchester, R. N., Lyne, A. G., D'Amico, N.,
Bailes, M., Gaensler, B. M., Nicastro, L., 1996,
MNRAS, 279, 1026

\reference
Johnston, S., Manchester, R. N., McConnell, D., Campbell-Wilson, D., 1999,
MNRAS, 302, 277

\reference
Kennel, C. F., Coroniti, F. V., 1984,
ApJ, 283, 710 

\reference
Kirk, J. G., Ball, L., Skj\ae raasen, O., 1999,
Astropart.\ Phys., 10, 31

\reference
Kirk, J. G., Ball, L., Skj\ae raasen, O., 2000,
ASP Conf., 202, 527

\reference
Melatos, A., Johnston, S., Melrose, D. B., 1995,
MNRAS, 285, 381 

\reference
Tavani, M., Arons, J., 1997,
ApJ, 477, 439

\reference
Yoshikoshi, T., et al., 1999,
Astroparticle Phys., 11, 267

\end{document}